\documentclass{article}
\usepackage{authblk}
\usepackage{amsmath, amssymb}
\usepackage{graphicx} 
\usepackage[utf8]{inputenc}
\usepackage{geometry}
\usepackage[T1]{fontenc}
\usepackage{mathptmx}
\usepackage{anyfontsize}
\usepackage{subcaption}
\usepackage{caption} 
\usepackage{authblk}
\DeclareCaptionLabelFormat{boldnumber}{\textbf{Table 1}}
\DeclareCaptionLabelSeparator*{spaced}{\\[1ex]}
\usepackage{changepage}

\usepackage[colorlinks=true, allcolors=blue]{hyperref}

\title{\textbf{\uppercase{The Energy Spectrum of the Pion from Lattice QCD}}}
\author[1]{Dilan Arik}
\author[2]{Navneeth C}
\author[3]{Sufiyan Mirza}
\author[4]{Ameer Mustafa}
\author[5]{Debarshi Mukherjee}
\author[6]{Rajath R}
\author[7]{Anantha Srinivasan}
\author[8]{Preksha Uniyal}
\author[9]{Martha Constantinou\thanks{Project Mentor for the REYES research program.}}

\affil[1]{Minerva University, 14 Mint Plaza, San Francisco, CA, 94103, USA}
\affil[2]{Department of Physics, Central University of Jharkhand, Ranchi, India - 835222}
\affil[3]{Department of Physics, Department of Mathematics, St. Xavier's College, Mumbai, India 400001}
\affil[4]{National University of Sciences and Technology (NUST), Department of Physics, School of Natural Sciences, Sector H-12, Islamabad, Pakistan 44000}
\affil[5]{Department of Physical Sciences, Indian Institute of Science Education and Research
Berhampur, Berhampur-760003, Odisha, India}
\affil[6]{Dept. of Physics, SPPU, Pune, India - 411007}
\affil[7]{Department of Physics, National Institute of Technology Calicut, Kozhikode, India - 673601
}
\affil[8]{Dept. of Physics \& Astrophysics, University Road, University Enclave, New Delhi-110007,India }
\affil[9]{Mentor: 
Temple University, 1801 N Broad St, Philadelphia, PA 19122, USA}

\date{}

\begin{document}

\maketitle
\begin{abstract}
    In this report, computational techniques are employed to extract the energy eigenvalues of the pion from two-point correlation function data that has been simulated using the lattice formulation of Quantum Chromodynamics (QCD) across various momentum values for the particle. The analysis focuses on systematically obtaining these eigenvalues to understand better the behavior of the pion under different kinematic conditions. Once extracted, these energy eigenvalues obtained through plateau fitting of the Energy vs Time graph, are utilized to plot the dispersion relation of the pion. This resulting dispersion relation is then compared with theoretical predictions to assess the accuracy and validity of the computational approach. The comparison provides insights into how well the lattice QCD simulations align with established theoretical expectations, and what range of kinematic variables are reliable.
\end{abstract}
\vspace{5mm}

\section{INTRODUCTION}
The Standard Model of particle physics is our most comprehensive theory for describing the forces in nature for fundamental particles and their interactions, excluding gravity. It explains quarks, leptons, gauge bosons, and the Higgs boson, which interact through electromagnetism, the weak and strong forces. Within the Standard Model, Quantum Chromodynamics (QCD) describes the strong interaction via a non-Abelian gauge theory with SU(3) symmetry. Quarks interact by exchanging gluons, which carry the colour charge, leading to complex interactions distinct from those of the electromagnetic force. For a detailed review, see \cite{Gross:2022hyw, Huston:2023ofk}.

Unlike other quantum field theories, such as Quantum Electrodynamics (QED), perturbative QCD fails at low energies because the physical coupling constants become large at that energy scale. Therefore, this warrants the use of non-perturbative approaches to study QCD effectively. One such approach is Lattice Quantum Chromodynamics. Lattice QCD involves discretizing space-time into a lattice grid, allowing for numerical simulations and calculations of physical observables, such as, in our case, the energy spectrum of fundamental particles. This method provides a way to study the properties of quarks and gluons in a controlled manner, overcoming the limitations of perturbative techniques. See (\cite{Lepage:1998dt}) for a brief review of Lattice QCD.

The most straightforward action for a gauge theory is given by the Yang-Mills action. Its lattice equivalent for a discretization of the QCD continuum action is the so-called Wilson's action (\cite{PhysRevD.10.2445}). However, several improved discretizations have been proposed. In this work, we use the \textit{Twisted Mass Fermion Action}, a modification of the \textit{Wilson Fermion action}. This modification improves control over discretization errors and offers automatic \(O(a)\) improvement (\cite{Frezzotti:2000}), although it still introduces explicit chiral symmetry breaking. However, the twisted mass formulation allows for better management of the chiral properties compared to the regular Wilson action, leading to more accurate results.

With a chosen lattice action in hand, one can compute several aspects of the QCD theory as in the continuum. The key aspect of the lattice formulation is the space-time discretization. Nevertheless, the complete QCD Lagrangian is simulated. Two-point functions are of particular interest since they provide a way to extract meaningful physical quantities, such as energy. The two-point function can be written in the spectral representation as (note that Wick rotation along the time axis has been performed),
\begin{equation}
    C(t) = \left\langle \hat{O}(t)\hat{O}^{\dagger}(0) \right\rangle = \sum_n A_n exp(-E_n t) = A_o exp(-E_0 t) \left(1+\frac{A_1}{A_0} exp(-(E_1-E_0) t)+\cdots \right)
\end{equation}
where $A_n$ is the matrix element $(A_n = |\left\langle n|\hat{O}|0 \right\rangle|^2)$. Energy can be extracted using the following relation,
\begin{equation} \label{eq:eff-energy}
    E_{eff} (t) = \log\left(\frac{C(t)}{C(t+1)}\right) \approx E_0 + (E_1 - E_0) \frac{A_1}{A_0} exp(-(E_1 - E_0)t) + \cdots
\end{equation}
As the time $t$ increases, the effective energy will approach the lowest energy state. This can be seen by the negative sign of the exponents that, for large $t$, give a smaller weight into the excited states, which have higher energy than the ground state.  In contrast, the effective energy will be a combination of various states for smaller values of $t$, and the ground state may be obtained via higher-order exponential corrections. The convergence to the lowest energy state will be faster if the matrix element $A_0$ is significantly greater than other matrix elements $A_i$ such that the ratio of the matrix elements in the sum is suppressed.

Using Monte Carlo simulations, the two-point correlation functions were computed on a lattice to extract crucial data for the analysis. These simulations were conducted on the \textbf{NERSC Perlmutter supercomputer}, and on the \textbf{OLCF SUMMIT supercomputer}, which provided the necessary computational power for handling the large datasets involved. The format of the dataset used for the analysis is outlined in Table~\ref{tab}.
\begin{table}[h!]
\caption{Statistics for the Pion Matrix Elements for Different Momentum Boosts}
\begin{center}
\renewcommand{\arraystretch}{1.7}
\begin{tabular}{l|cccccc}
\hline
$P_3$ [GeV] & $\quad$0$\quad$ & $\quad$$\pm$0.41$\quad$ & $\quad$$\pm$0.83$\quad$ & $\quad$$\pm$1.24$\quad$ & $\quad$$\pm$1.66$\quad$ & $\quad$$\pm$2.07$\quad$  \\ \hline
$N_{\rm confs}$ & 1,198 & 1,198 & 1,198 & 1,198 & 1,198 & 1,198 \\\hline
$N_{\rm src}$ & 1 &  8 &  8 &  8 &  24 & 200  \\\hline
$N_{\rm tot}$ & 1,198 & 9,584 & 9,584 & 9,584  & 28,752  & 239,600 \\
\hline
\end{tabular}
\end{center}
\textit{Note:} $N_{\rm confs}$, $N_{\rm src}$, $N_{\rm total}$ are the number of configurations, source positions per configuration, and total statistics, respectively.
\label{tab}
\end{table}

In the following sections, the report will detail the procedures for performing plateau fitting of the effective energy versus time graph and the analysis of the dispersion relation graph.
\section{METHODS}
\subsection{DATA PREPARATION}
The data sets utilized in this research work consist of the two-point correlation functions at discrete Euclidean time slices, and the configurations are basically snapshots of the QCD vacuum. Redefining the time to Euclidean time allows us to use statistical methods to calculate the expectation values of the operators. For each configuration, we obtain the two-point correlation function of the pion with momentum-boosted values, $aP$, ranging from 0 to $a|P|=\pm 10\pi/L$ that is about 2 GeV. These snapshots enable observation of the evolution of physical quantities.

The data contain complex valued correlation functions, of which only the real parts were significant for our analysis, as the energy is a real quantity. As shown in Table~\ref{tab}, a significant number of configurations are utilized, ${\cal O}(1000)$. For mesons, we also increase the statistics by averaging on $\pm P$ values and on time due to the time-reversal symmetry to reduce the statistical uncertainties due to gauge noise.

\subsection{JACKKNIFE RESAMPLING}
The jackknife error analysis method is particularly useful in scenarios where the data exhibits correlations, providing accurate estimates of statistical errors (\cite{Sinharay:2010}).

This method works by generating multiple subsets of the original data, in our case, 1198, where each subset is created by omitting one data point (in our case, configuration) at a time. This process is executed for each Euclidean time slices ranging from \(0\) to \(63\). For each of these subsets, the effective energy is calculated. This process yields a series of estimates, which are then used to determine the mean and the variance. The jackknife error is particularly useful because it provides an unbiased estimate of the statistical error.

\subsection{EXTRACTION OF THE EFFECTIVE ENERGY}
The effective energy is determined by analyzing the jackknife resampled two-point correlation
functions (bins) using equation (\ref{eq:eff-energy}).
This method is applied consistently across momentum-boosted states in multiple directions to provide a comprehensive understanding of the pion’s behavior as a function of $t$. A representative example can be seen in the left panel of Figure~\ref{fig:graph1}, which corresponds to a momentum boost of $a|P|=10\pi/L$, which in physical units is about 2 GeV. It is worth mentioning that the statistics for this case is 239,600 data. The plot shows the effective energy as a function of $t$. Such plots identify a region of $t$, where a ground state appears as a plateau. 

\subsection{PLATEAU FITTING}
Once the central energy values have been obtained and plotted against time, it can be observed that as time increases, the energy values level out, reaching a plateau. Using this plateau, the ground state energy must be extracted.

The energy is extracted using model fitting with least squares. The energy is modelled as a constant, i.e. the variations of the energy values are considered as random variations about a constant mean value in the region in which the plateau is observed. If the energy value at time \(t\) is \(E \left( t \right)\) and errors in the energy values are \(\sigma_E \left( t \right)\), the value for the mean value in the plateau region is given by:
\begin{equation}
  E_{dis} = \frac{\sum_{t = t_1}^{t = t_2} E \left( t \right)/ \sigma_E \left( t \right)^2}{\sum_{t = t_1}^{t = t_2} 1 / \sigma_E \left( t \right)^2}    
\end{equation}
where the plateau is fitted in the region \(t \in \left[ t_{1}, t_{2} \right]\).

This process is carried out on each bin individually before the calculation of the central values for the energy. Thus, an \(E_{dis}\) is obtained for each bin, for which the jackknife analysis is performed again before obtaining a single central value.
\subsection{Dispersion Relation}
After obtaining the plateau fitted values for the energy \(E_{dis}\), this is plotted and compared  against the theoretical values obtained from the dispersion relation (Figure \ref{fig:graph2}). The lattice data are compared with theoretical predictions. In particular, using the special theory of relativity, we know that the energy of a massive particle is found using the following relation:
\begin{equation}
E = \sqrt{m^2 + \mathbf{p}^2}
\end{equation}
where \(m\) is the rest mass of the particle and \(\mathbf{p}\) is the momentum boost of the particle (the speed of light has been set to \(1\) here). Note that \(\mathbf{p}\) is a vector and has components \(p_{x}\), \(p_{y}\), and \(p_{z}\).
As mentioned before, the time \(t\) is discretized in lattice QCD calculations. As a result, the energy values are also discretized leading to:
\begin{equation}
aE = \sqrt{\left(am\right)^2 + a^2\mathbf{p}^2}
\end{equation}
\section{RESULTS AND DISCUSSION}
\par Here, we show the results. The first part of Figure 1 is dedicated to an example of plateau fitting and the central mean values for effective energy as explained in the previous section. The second graph depicts the theoretical dispersion relation with the simulation results. 

\begin{figure}[h]
    \centering
    \begin{subfigure}[b]{0.45\textwidth}
        \includegraphics[width=\textwidth]{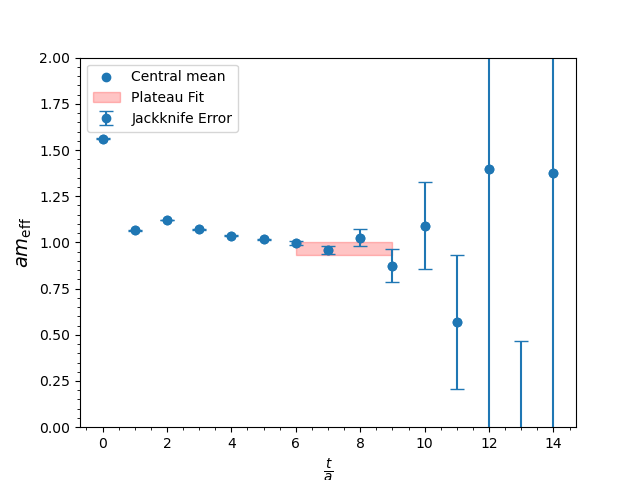}
        \caption{Central Mean Values and Plateau Fitting.}
        \label{fig:graph1}
    \end{subfigure}
    \begin{subfigure}[b]{0.45\textwidth}
        \includegraphics[width=\textwidth]{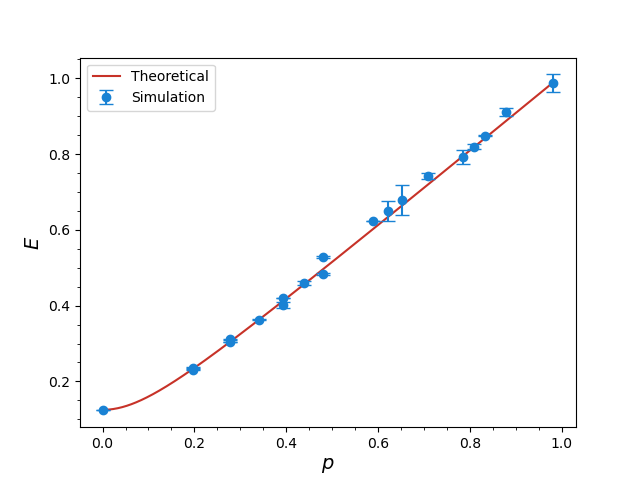}
        \caption{Dispersion Relation.}
        \label{fig:graph2}
    \end{subfigure}
    \caption{{Energy for the pion.  The first graph depicts the effective energy/mass values for the pion with jackknife errors. We observe that the jackknife errors significantly increase with time and that effective energy decay, allowing us to find the plateau region. The red band shows where there is a plateau. The second graph compares the theoretical dispersion relation and our simulation results. Note that all momentum values are in GeV/c and a=0.0934 fm. }}
    \label{fig:side_by_side}
\end{figure}

As can be seen from Figure~\ref{fig:graph1}, the data show the clear decay of energy values to the ground state. This is expected as the exponential terms for excited states and multi-hadron states become very small, indicating that the contributions from these states diminish and the ground state becomes predominant. Another trend seen as time increases is that the noise-to-signal ratio increases, but as we only need the plateau for our purposes, this increase is inconsequential. The plateau fits also show the expected rise in ground state energy with the increase in momentum. These results also clearly indicate the increase in error at higher momentum values; this necessitates the increased statistics as indicated in Table 1.

The trend of the plateau occurring at lower time slice values is justified by the fact that the excited states decay faster. Another thing to notice is the decrease in the length of the acceptable plateau region; this is a clear indication of the difficulty that arises in analyzing higher momenta if the ground state lasts for a smaller and smaller amount of time, it creates uncertainty as to if this is actually the ground state. This is one of the main sources of errors that were encountered in the analysis of higher momentum values.

As seen in Figure~\ref{fig:graph2}, the lattice QCD data are in excellent agreement with the continuum dispersion relation; almost all points lie either on the graph or within the given error values. However, one to two points demonstrate a small difference from the theoretical line. This is justified by the facts mentioned in the previous subsection, about the problems that arise with analysis of higher momentum values, the statistics in this regime become unreliable and in turn, so do their errors. 

\section{CONCLUSION}
\par In conclusion, we have shown how closely the simulation results for the pion match the theoretical dispersion relation through the analysis of the 2-point function results using Jackknife Resampling and plateau fitting. This highlights the importance of the non-perturbative approaches of lattice QCD for explaining the effects of the strong force and differentiating between signal and noise. We can explore higher momentum states for future work, allowing us to observe how pion behaves in these higher momentum-boosted ensembles. We should note that the data gets noisier as we move into the higher momentum domain, so this might be difficult to capture fully. We can also explore different lattice volumes and compare how our results might scale with the lattice volume. Lastly, we can consider repeating the calculations on ensembles using smaller lattice spacing, allowing us to get closer to the continuous reality that we have omitted by the discretization of space and time.

\section{\textbf{ACKNOWLEDGEMENTS}}





We are immensely grateful to the coordinators of the REYES Mentorship Program for providing us the opportunity to work with experts and enhance our research skills. We extend our appreciation to Professor Dr. Martha Constantinou, who
mentored us during the program. We would also like to thank Josh Miller and Isaac Anderson for valuable discussions on this subject. Lastly, we would like to acknowledge the rest of the members in our collaboration. In no particular order: Rahul V. Silva, Shaurya Malhotra, Prashant Banerjee, Aadvait Pillai, Manas Agrawal, Shibashis Mukhopadhyay, Anirudh J, Ashwath Raman, Mayul Verma, Sukriti Atrey, MD Kaish, Sai Krishna, Shaikha Albassam, Phillip Crosby, Giovanni, Marie, Janna, Zain Mothupi, David Delgado, Dario Roman Gomez Martin, William Dang, Sadeed Anwar.
M.C. acknowledges financial support from the U.S. Department of Energy, Office of Nuclear Physics, Early Career Award under Grant No.\ DE-SC0020405, and Grant No.\ DE-SC0025218.
Interactions within the Quark-Gluon Tomography (QGT) Topical Collaboration, funded by the U.S. Department of Energy, Office of Science, Office of Nuclear Physics, with Award DE-SC0023646, have benefited this work. 
This research used resources of the National Energy Research Scientific Computing Center, a DOE Office of Science User Facility supported by the Office of Science of the U.S. Department of Energy under Contract No. DE-AC02-05CH11231 using NERSC award NP-ERCAP0027642, as well as NERSC award ALCC-ERCAP0030652. Also, some of the data used in this work have been produced using a computational award from Oak Ridge Leadership Computing Facility through the Summit Plus program (project: NPH160).

\end{document}